\documentstyle[12pt]{article}
\input{epsf}
\def\O16{^{16}{\rm O}}

\begin{document}
\def\title#1{\begin{center}{\Large #1}\end{center}}
\def\author#1{\vspace*{0.5ex}\begin{center}{#1}\end{center}}
\def\address#1{\begin{center}\vspace*{-1ex}{\it #1}\end{center}}  
\def\pubnum{329/NP}
\def\mark#1{\vspace*{1cm} \fbox{ #1}}
\def\NOTE#1{{\bf (*#1*)}}
\def\Isom#1{{\mbox{{\rm Isom}$#1$}}}
\def\UC#1{\widetilde #1}
\def\ME#1{#1'}
\begin{titlepage}
\hfill
\parbox{6cm}{{TIT/HEP-\pubnum} \par May,1996}
\par
\vspace{21mm}
\title{\bf Nuclear G-Matrix Elements from Nonlocal Potentials}
\author{
K.~Yoshida, A.~Hosaka$^{(a)}$, and M.~Oka\\[0.5cm]
Department of Physics, Tokyo Institute of Technology\\
Meguro, Tokyo 152, Japan\\[0.5cm]
$^{(a)}$Numazu College of Technology\\
3600 Ooka, Numazu 410, Japan\\[0.5cm]
e-mail: kyoshida@th.phys.titech.ac.jp\\
hosaka@la.numazu-ct.ac.jp\\
oka@th.phys.titech.ac.jp
}

\begin{abstract}
We study effects of nonlocality in the nuclear force on the G-matrix
elements for finite nuclei.
Nuclear G-matrix elements for $\O16$ are calculated in the harmonic
oscillator basis
from a nonlocal potential which models quark exchange
effects between two nucleons.
We employ a simple form of potential that
gives the same phase shifts as a realistic local nucleon potential.
The G-matrix elements calculated from the nonlocal potential show
moderate increase in repulsion
from those derived from the local potential.

\end{abstract}

\end{titlepage}

\section{Introduction}

Several models of the nucleon-nucleon interaction have been
presented and applied to nuclear physics problems.
The G-matrix interaction
plays a central role in calculations of
nuclear structure from the nucleon-nucleon ($NN$) interaction,
as it is the basic two-body interaction in many body systems.

It is well known that the nucleon-nucleon interaction is
described by the quark
cluster model (QCM).
This model gives a good description
of the nucleon-nucleon scattering ~\cite{QCM1,QCM2}.
The main feature of
the interaction is nonlocality at short distances.
This short range potential has a nonlocal Gaussian soft
core of the form,
\[ V({\bf r'},{\bf r})\propto\exp [-(\frac{{\bf r}-{\bf
r}'}{b})^2]. \]
This type of nonlocality is expected from the quark
antisymmetrization between the baryons.
The Gaussian form reflects the quark wave function inside the baryon
and the nonlocality range parameter $b$ is determined by the size of
the baryon.
The nonlocal exchange force can reproduce the short-range repulsion of
the nuclear force at low energy.
The repulsive ``core'' in QCM is a soft core with energy dependence in
its equivalent local form.  It allows the $NN$ relative wave function
to go inside and thus gives a milder form factor of the deuteron (or
nuclei) at large momenta ~\cite{QCM3}.
Because the quark exchange nonlocality comes only at short distances,
the central part of the potential is most affected.  In fact, the $NN$
repulsion is roughly spin-isospin independent.

In this paper, we would like to study significance of the nonlocality
of quark model origin in nuclear structure calculations.  
Since we can not see signals of such nonlocality in on-shell properties,  
it would be interesting if we could see them in 
nuclear phenomena.  
To do so, we calculate G-matrix elements from nonlocal potentials.  
In conventional nuclear physics, the G-matrix interaction is used as 
the basic two-body interaction for the nuclear many-body problem.  
Hence, if there would be signals of quark-originated nonlocality in nuclear 
phenomena, they must exist in the G-matrix interaction as well.  
Previously, effects of symmetry restoration of QCD in G-matrix elements 
were examined~\cite{HoTo}.  
Thus in the present study we investigate another possibility of 
QCD oriented effects in nuclear physics.  

This paper is organized as follows.  
In section 2, a
method to calculate G-matrix elements for finite
nuclei is given briefly.
In section 3, we
introduce the Tamagaki G3RS (Gaussian soft core potential with three ranges)
potential~\cite{TAMA} as a local potential.
Then we construct a nonlocal Gaussian soft core potentials,
which has the same intermediate and long range parts as the original
local potential.
The short-range part is replaced by a nonlocal potential with free
parameters which are determined so as to reproduce the same scattering phase
shifts as the local potential.
In section 4, numerical results for the
G-matrix elements are given both for the local and nonlocal
potentials.
A summary is given in section 5.

\section{Nuclear G-matrix for Finite Nuclei}

The G-matrix is the most fundamental two body interaction which
incorporates the simplest many body effect, i.e., the Pauli
blocking.
It is obtained by solving the Bethe-Goldstone equation~\cite{BETHE}:
\begin{equation}
G(E)=v+v\frac{Q}{E-H_0}G(E).
\label{eq:BG}
\end{equation}
Here $v$ is a nucleon-nucleon interaction
in free space, $Q$ the Pauli-blocking
operator, and $H_0$ a single particle hamiltonian.
We compute G-matrix elements for finite nuclei,
specifically for $\O16$, following
the method developed in previous works~\cite{BBH,ATB,HKT}.
$H_0$ is chosen to be the harmonic oscillator hamiltonian.
The two-body problem is then reduced to a one-body problem of the relative
coordinate by employing the Eden and Emery's approximation ~\cite{EE}
for the $Q$ operator.
Explicitly, after reducing the Pauli-blocking operator as a function of
the relative coordinate, it is assumed to be
\begin{equation}
\label{Q16}
Q(n^\prime) = \left\{ \begin{array}{c c}
                      1 & {\rm for} \; n^\prime > n+2 \\
                      0 & {\rm otherwise}
                      \end{array}
                      \right.
\end{equation}
where $n$ is the node quantum number for the relative motion of the
initial two nucleons and $n^\prime$ that for intermediate states.
The Barret-Hewitt-MacCarthy's method ~\cite{BHM} is then used
for matrix inversion.
Additional parameter is introduced for the gap energy, which is chosen to
be 40 MeV to achieve a good agreement between the calculated G-matrix
elements and empirical ones~\cite{Abrown}.
Finally, the starting energy is averaged over the
Fermi surface by assuming that the relevant two nucleons are there.

G-matrix elements are obtained first in partial wave channels
denoted by $^1S_0$, $^1P_1$, etc.,  which are then transformed into those in
interaction channels denoted by SE (Singlet Even), TE (Triplet Even), etc.
Matrix elements in SE and SO (Singlet Odd) channels are just those in
$^1S_0$ and $^1P_1$ channels, respectively.
The TE and TNE (Tensor Even) components are obtained from the coupled
$^3S_1-^3D_1$ channels:
\begin{eqnarray}
G({\rm TE}) &=& \langle S \mid G(^3S_1-^3D_1)\mid S \rangle, \\
G({\rm TNE}) &=& \langle S \mid G(^3S_1-^3D_1)\mid D \rangle.
\end{eqnarray}
The TO, TNO, LSO (LS Odd) and LSE (LS Even) 
components are obtained from the following relations~\cite{BBH}:
\begin{eqnarray}
G({\rm TO}) &=& G(^3P_0)+2G(LSO)+4G(TNO),\\
G({\rm TNO}) &=& -\frac{5}{72}[ 2G(^3P_0)-3G(^3P_1)+G(^3P_2) ], \\
G({\rm LSO}) &=& -\frac{1}{12}[ 2G(^3P_0)+3G(^3P_1)-5G(^3P_2) ], \\
G({\rm LSE}) &=& -\frac{1}{60}[ 9G(^3D_1)+5G(^3D_2)-14G(^3D_3) ],
\end{eqnarray}

In the previous studies~\cite{BBH,ATB,HKT},  the Reid soft-core
potential, Paris potential, and other phenomenological nuclear forces
have been applied.
We here choose the Tamagaki G3RS potential~\cite{TAMA},
which is a three range gaussian with a short range soft core.
We replace the shortest-range part of the original potential by
a nonlocal one,
and apply it to the G-matrix calculation.
The resulting  G-matrix elements are compared with each other.

\section{Gaussian Nonlocal Potential}

We use the following Tamagaki potential as a standard local potential,
\begin{equation}
V_L({\bf r})=V_C(r)+S_{12}V_T(r)+({\bf L}\cdot{\bf
S})V_{LS}(r)+W_{12}V_W(r)+{\bf L}^2V_{LL}(r),
\label{eq:tam}
\end{equation}
with
\begin{eqnarray*}
 S_{12}&=&3(\sigma_1\cdot {\bf r})(\sigma_2\cdot {\bf
r})/r^2-\sigma_1\cdot\sigma_2 \\
 W_{12} &=& 1/2\{ ({\bf\sigma}_1\cdot {\bf L})({\bf\sigma}_2\cdot {\bf L})
 +({\bf\sigma}_2\cdot {\bf L})({\bf\sigma}_1\cdot {\bf L}) \}
 -({\bf\sigma}_1\cdot{\bf\sigma}_2){\bf L}^2/3 \\
 &=& {\bf (L\cdot S)}^2 
-\{ \delta_{LJ}+({\bf\sigma}_1\cdot{\bf\sigma}_2)/3 \} {\bf L}^2 
\end{eqnarray*}
where $L$ and $J$ stand for the orbital and total angular momenta,
respectively.
The radial functions $V_n(r)$ ($n=C$, $LS$, \ldots) are given by
\begin{equation}
V_n(r)=\sum_{i=1}^3 V_{ni}\exp[-(r/\eta_{ni})^2],
\label{eq:tamrad}
\end{equation}
where the suffices $i=1,2,3$ refer to the short range, intermediate
range and long range part, respectively.
The potential parameters, $ V_{ni}$ and $\eta_{ni}$, are determined so
as to reproduce the scattering phase shifts  
as given in Table \ref{tab:g3rs} ~\cite{TAMA}.

The nonlocal potential is constructed by replacing the short range
part ($i=1$) of Eq.~(\ref{eq:tamrad}).
We consider nonlocality only for the central part of the form:
\begin{equation}
V({\bf r'},{\bf r})= V_{NL}({\pi}^{1/2}b)^{-3}
\exp [-(\frac{{\bf r}-{\bf r}'}{b})^2-
(\frac{{\bf r}+{\bf r'}}{\eta_{NL}})^2]
\label{eq:nlpot}
\end{equation}
Similar nonlocal terms might appear in noncentral forces,
but for the reason explained in introduction, we consider the nonlocality only 
in the central potential.
In Eq.~(\ref{eq:nlpot}), the factor
$({\pi}^{1/2}b)^{-3} \exp [- (\frac{{\bf r}-{\bf r'}}{b})^2]$ represents
the nonlocality of the nuclear force at short distances, with a new
parameter $b$ as the range of nonlocality.
When $b\rightarrow 0$, this part
reduces to the delta function,
$\delta({\bf r}-{\bf r'})$,
and then the nonlocal potential becomes the local one.
In the quark cluster model, $b$ is proportional to the size of the quark
wave function in the nucleon.
A typical value of $b$ would be 0.5 fm, which we use in the following 
calculations.

In order to study effects of nonlocality exclusively, we
require the nonlocal potential to reproduce the same
scattering phase shifts
as the corresponding local potential does.  
We call such a nonlocal potential as local-equivalent (LE) nonlocal 
potential.  
For this, we adjust the potential parameters, $V_{NL}$ and
$\eta_{NL}$.
We allow them to depend on the angular momentum $L$.
Phase shifts in
various partial waves at scattering energies from
$E_{cm} = 0$ to $300$ MeV are fitted.
As shown in Figs. 1 - 2, both the local and LE nonlocal potentials (LENL)
reproduce empirical phase shifts well up to the scattering energy $\sim$
300 MeV.
In particular, the difference between the local and LENL are
negligibly small.
In Figs.~1 and 2, we also present the phase shifts calculated
from the nonlocal potentials (NL) with the same strength,
$V_{NL}=V_{C1}$ and the same range $\eta_{NL}=\eta_{C1}$
as the original Tamagaki potential.
They show that the nonlocality yields a softer repulsion at large
energy.
Although we have shown here the phase shifts only for the two channels, we
find qualitatively the same feature for other $S$ and $P$ channels.

Summarizing, our LE nonlocal potential reads
\begin{eqnarray}
V_{NL}({\bf r'},{\bf r})&=&V_{NL}({\pi}^{1/2}b)^{-3}
\exp [-(\frac{{\bf r}-{\bf r}'}{b})^2-
(\frac{{\bf r}+{\bf r'}}{2\eta_{NL}})^2]\nonumber \\
&&+\sum_{i=2}^3 V_{Ci} \exp[-(r/\eta_{Ci})^2]\, \delta({\bf r}-{\bf r}')
\label{eq:mnlp}
\end{eqnarray}
with the parameters given in Table \ref{tab:g3rs} for the local part and 
in Table \ref{tab:le0.5} for the nonlocal part ($b=0.5$ fm).

\begin{table}
 \doublerulesep=0pt
 \begin{center}
  \begin{tabular}{@{\extracolsep{\fill}}|c|ccc|ccc|cc|c|c|}\hline
         &$\eta_{C_1}$&$\eta_{C_2}$&$\eta_{C_3}$&$\eta_{T_1}$&
            $\eta_{T_2}$&$\eta_{T_3}$&$
         \eta_{LS_2}$&$\eta_{LS_3}$&$\eta_{W_2}$&$\eta_{LL_2}$ \\
         &$V_{C_1}$&$V_{C_2}$&$V_{C_3}$&$V_{T_1}$&$V_{T_2}$&$V_{T_3}$&
           $V_{LS_2}$&$V_{LS_3}$&
         $V_{W_2}$&$V_{LL_2}$ \\ \hline
   $SE$ &0.447&0.942&2.5& & & & & & &0.942 \\
                 &2000&-270&-5& & & & & & &15       \\ \hline
   $TO$ &0.447&0.942&2.5&0.447&1.2&2.5&0.447&0.6&0.942&0.942 \\
   &2500&-70&1.67&-20&20&2.5&600&-1050&0&0      \\ \hline
   $SO$ &0.447&0.942&2.5& & & & & & &0.942 \\
         &2000&50&10&&&&&&&0 \\ \hline
   $TE$ &0.447&0.942&2.5&0.447&1.2&2.5&0.447&0.6&0.942&0.942 \\
         &2000&-230&-5&67.5&-67.5&-7.5&0&0&-30&30       \\ \hline
  \end{tabular}
 \end{center}
 \caption{Parameters of the Tamagaki G3RS potential}
 \label{tab:g3rs}
\end{table}

\begin{table}
  \begin{center}
   \begin{tabular}{|r|c|c|}\hline
                &$\eta_{NL}^{b=0.5}$            &$V_{NL}^{b=0.5}$ \\ \hline
        $l=0$   &$0.894\times \eta_{C_1}$       &$1.45\times V_{C_1}$ \\
        $1$     &$0.894\times \eta_{C_1}$       &$2.9\times V_{C_1}$ \\
        $2$     &$0.894\times \eta_{C_1}$       &$6.0\times V_{C_1}$ \\
        $3$     &$0.894\times \eta_{C_1}$       &$12.5\times V_{C_1}$ \\
        $4$     &$0.894\times \eta_{C_1}$       &$27\times V_{C_1}$ \\ \hline
 \end{tabular}
 \end{center}
 \caption{Parameters of the nonlocal  potential with $b=0.5$ fm}
 \label{tab:le0.5}
\end{table}

\section{Numerical Results}

The G-matrix elements in the harmonic-oscillator basis are presented in
Tables \ref{tab:gl}--\ref{tab:gle} 
for the local (Tamagaki) and LE
nonlocal potentials with $b$ = 0.5 fm
as functions of node quantum numbers $n$ and $n'$.
These are calculated using an oscillator parameter $\hbar\omega=14$ MeV
for $\O16$ and with the gap energy of 40 MeV, which
are the same as those used in Refs.~\cite{BBH,ATB,HKT}.

\begin{table}
 \doublerulesep=0pt
 \begin{center}
  \begin{tabular*}{14cm}{@{\extracolsep{\fill}}cr|cccc} 
    \hline\hline
                &s s    &n=0    &n=1    &n=2    &n=3    \\ \hline
singlet         &n'=0   &-6.605 &-5.182 &-3.551 &-2.064 \\
even            &1      &       &-4.564 &-3.251 &-1.831 \\
                &2      &       &       &-2.352 &-1.233 \\
                &3      &       &       &       &-0.428 \\ \hline\hline
                &s s    &n=0    &n=1    &n=2    &n=3    \\ \hline 
triplet         &n'=0   &-10.495&-8.725 &-6.488 &-4.401 \\
even            &1      &       &-7.982 &-6.209 &-4.274 \\
                &2      &       &       &-5.028 &-3.525 \\
                &3      &       &       &       &-2.468 \\ \hline\hline
                &p p    &n=0    &n=1    &n=2    &n=3    \\ \hline
singlet         &n'=0   &2.365  &2.140  &1.768  &1.495  \\
odd             &1      &       &2.614  &2.575  &2.374  \\
                &2      &       &       &2.905  &2.935  \\
                &3      &       &       &       &3.202  \\ \hline\hline
                &p p    &n=0    &n=1    &n=2    &n=3    \\ \hline 
triplet         &n'=0   &0.200  &0.057  &-0.052 &-0.098 \\
odd             &1      &       &0.039  &-0.005 &-0.031 \\
                &2      &       &       &0.030  &0.059  \\
                &3      &       &       &       &0.144  \\ \hline\hline
                &s d    &n=0    &n=1    &n=2    &n=3    \\ \hline
tensor          &n'=0   &-5.391 &-7.403 &-8.463 &-9.011 \\
even            &1      &-2.429 &-4.688 &-6.463 &-7.691 \\
                &2      &-0.967 &-2.449 &-4.067 &-5.502 \\
                &3      &-0.338 &-1.095 &-2.208 &-3.454 \\ \hline\hline
                &p p    &n=0    &n=1    &n=2    &n=3    \\ \hline
tensor          &n'=0   &0.772  &0.745  &0.639  &0.535  \\
odd             &1      &       &0.896  &0.872  &0.782  \\
                &2      &       &       &0.940  &0.908  \\
                &3      &       &       &       &0.939  \\ \hline\hline
                &d d    &n=0    &n=1    &n=2    &n=3    \\ \hline
LS              &n'=0   &-0.034 &-0.048 &-0.057 &-0.062 \\
even            &1      &       &-0.075 &-0.091 &-0.102 \\
                &2      &       &       &-0.117 &-0.133 \\
                &3      &       &       &       &-0.156 \\ \hline\hline
                &p p    &n=0    &n=1    &n=2    &n=3    \\ \hline
LS              &n'=0   &-0.416 &-0.666 &-0.850 &-0.983 \\
odd             &1      &       &-0.998 &-1.247 &-1.432 \\
                &2      &       &       &-1.536 &-1.753 \\
                &3      &       &       &       &-1.990 \\ \hline\hline
  \end{tabular*}
 \end{center}
 \caption{G-matrix elements from the local potential for $\O16$}
\label{tab:gl}
\end{table}

\begin{table}
 \doublerulesep=0pt
 \begin{center}
  \begin{tabular*}{14cm}{@{\extracolsep{\fill}}cr|cccc} 
    \hline\hline
                &s s    &n=0    &n=1    &n=2    &n=3    \\ \hline
singlet         &n'=0   &-6.542 &-5.098 &-3.444 &-1.931 \\
even            &1      &       &-4.460 &-3.126 &-1.683 \\
                &2      &       &       &-2.213 &-1.076 \\
                &3      &       &       &       &-0.260 \\ \hline\hline
                &s s    &n=0    &n=1    &n=2    &n=3    \\ \hline
triplet         &n'=0   &-10.499&-8.714 &-6.457 &-4.343 \\
even            &1      &       &-7.962 &-6.171 &-4.212 \\
                &2      &       &       &-4.978 &-3.458 \\
                &3      &       &       &       &-2.392 \\ \hline\hline
                &p p    &n=0    &n=1    &n=2    &n=3    \\ \hline
singlet         &n'=0   &2.372  &2.150  &1.780  &1.507  \\
odd             &1      &       &2.627  &2.591  &2.392  \\
                &2      &       &       &2.924  &2.955  \\
                &3      &       &       &       &3.224  \\ \hline\hline
                &p p    &n=0    &n=1    &n=2    &n=3    \\ \hline
triplet         &n'=0   &0.206  &0.064  &-0.046 &-0.093 \\
odd             &1      &       &0.046  &0.000  &-0.028 \\
                &2      &       &       &0.032  &0.056  \\
                &3      &       &       &       &0.134  \\ \hline\hline
                &s d    &n=0    &n=1    &n=2    &n=3    \\ \hline
tensor          &n'=0   &-5.400 &-7.421 &-8.491 &-9.048 \\
even            &1      &-2.439 &-4.707 &-6.493 &-7.732 \\
                &2      &-0.978 &-2.469 &-4.097 &-5.543 \\
                &3      &-0.348 &-1.114 &-2.236 &-3.493 \\ \hline\hline
                &p p    &n=0    &n=1    &n=2    &n=3    \\ \hline
tensor          &n'=0   &0.772  &0.745  &0.639  &0.537  \\
odd             &1      &       &0.897  &0.873  &0.784  \\
                &2      &       &       &0.942  &0.912  \\
                &3      &       &       &       &0.944  \\ \hline\hline
                &d d    &n=0    &n=1    &n=2    &n=3    \\ \hline
LS              &n'=0   &-0.034 &-0.048 &-0.057 &-0.062 \\
even            &1      &       &-0.075 &-0.091 &-0.102 \\
                &2      &       &       &-0.117 &-0.134 \\
                &3      &       &       &       &-0.157 \\ \hline\hline
                &p p    &n=0    &n=1    &n=2    &n=3    \\ \hline
LS              &n'=0   &-0.414 &-0.663 &-0.849 &-0.983 \\
odd             &1      &       &-0.997 &-1.248 &-1.436 \\
                &2      &       &       &-1.541 &-1.762 \\
                &3      &       &       &       &-2.006 \\ \hline\hline
  \end{tabular*}
 \end{center}
 \caption{ G-matrix elememnts from LE nonlocal potential (b=0.5fm) for $\O16$}
\label{tab:gle}
\end{table}

First we compare G-matrix elements calculated from various nuclear forces.
For this purpose, we show in Table \ref{tab:gcom}
the G-matrix elements derived from the Paris
potential and those of the present calculation for the SE channel.
It turns out that both matrix elements are very similar.
The difference is significant only for less important off-diagonal matrix 
elements, which is, however, typically 20 \% level.  
This is interesting because the Tamagaki potential has the Gaussian
tail and its functional form differs from the others.
This fact suggests that G-matrix elements are strongly constrained by
on-shell properties (phase shifts).  
We have confirmed this by calculating G-matrix elements using 
$NN$ potentials which do not necessarily
reproduce empirical phase shifts.  

\begin{table}
\begin{center}
        \begin{tabular}{c c| c c c c}
                \hline
              & $n^\prime$ & $n = 0$ & $n = 1$  &  $n = 2$ & $n = 3$  \\
        \hline
       	  & 0 & -6.605 & -5.182 & -3.551 & -2.064 \\
 Tamagaki & 1 &        & -4.564 & -3.251 & -1.831 \\
          & 2 &        &        & -2.352 & -1.233 \\
          & 3 &        &        &        & -0.428 \\
        \hline
          & 0 & -6.580 & -5.292 & -3.743 & -2.502  \\
 Paris    & 1 &        & -4.644 & -3.322 & -1.947  \\
          & 2 &        &        & -2.386 & -1.070  \\
          & 3 &        &        &        & -0.318  \\
                \hline
        \end{tabular}
        \end{center}
        \caption{Comparison of diagonal G-matrix elements in the SE channel
        derived from three nuclear forces.}
	\label{tab:gcom}
\end{table}

Now we turn to see the effects of nonlocality
in the G-matrix elements.
From Tables \ref{tab:gl}-\ref{tab:gle},
one sees that for non-central channels of tensor and LS, the
results of the local and of LE nonlocal potentials are almost
identical.
This seems natural because in the
present calculation nonlocality is introduced only in the central
channel.  

Effects of nonlocality in the G-matrix elements are best seen in the
S-wave channels of SE and TE.
They are repulsive, though the effects are typically as small as a few \%
or even less.  
Therefore, one may conclude that 
the present nonlocality from the quark exchange effects are negligibly 
small as compared with other many-body effects of nuclear physics which is 
typically about 10 \% or even more~\cite{Abrown}.  
The reason that the present nonlocality acts as repulsive is understood
in the following way.
As we have discussed in section 3, 
when nonlocality is introduced for a repulsive component,
it effectively reduces the repulsion in the $NN$ potential.
To get the local-equivalent phase shifts, one has to modify the potential
parameters by reducing the range and increasing the strength of repulsion
as summarized in Table \ref{tab:le0.5}.
In the calculation of the G-matrix elements, the Pauli-blocking operator 
$Q$ 
forbids the transition to lower levels which are already occupied.
This effectively reduces the chance for the two nucleons to feel
attraction in the nuclear force
which comes mainly from the transition to the lower states.
Hence, in the G-matrix elements, repulsive components are more enhanced.  

The nonlocal effect would become important for heavier nuclei where more
levels are Pauli blocked.
We have performed a calculation of G-matrix elements in
the zirconium region.
The calculation was done  in the same way as for $\O16$, but with
$\hbar \omega$ = 8.8 MeV and with the following Pauli-blocking operator:
\begin{equation}
Q(n^\prime) = \left\{ \begin{array}{c c}
                      1 & {\rm for} \; n^\prime > n+4 \, ,\\
                      0 & {\rm otherwise} \, .
                      \end{array}
                      \right.
\end{equation}
Here, as the mass number is increased, more states are Pauli-blocked, 
which is implemented in the inequality $n^\prime > n+4$ in place of 
$n^\prime > n+2$ for $\O16$ in (\ref{Q16}).
Results are summarized in Table \ref{tab:gzr} for the SE channel,
where we see slightly more nonlocal effects.  
The difference is, however, once again very small.

\begin{table}
\begin{center}
        \begin{tabular}{c c| c c c c}
                \hline
              & $n^\prime$ & $n = 0$ & $n = 1$  &  $n = 2$ & $n = 3$  \\
        \hline
	  & 0 & -3.922 & -3.551 & -2.926 & -2.291 \\
local	  & 1 &        & -3.449 & -2.961 & -2.360 \\
          & 2 &        &        & -2.612 & -2.114 \\
          & 3 &        &	&        & -1.718 \\
        \hline
            & 0 & -3.812 & -3.422	& -2.787 & -2.145 \\
LE nonlocal & 1 &        & -3.299 & -2.800 & -2.191 \\
            & 2 &        &        & -2.440 & -1.934 \\
            & 3 &        &        &        & -1.533 \\
                \hline
        \end{tabular}
        \end{center}
        \caption{G-matrix elements in the SE channel for zirconium }
\label{tab:gzr}
\end{table}

\section{Summary}

The G-matrix elements for $\O16$ have been calculated in the harmonic
oscillator basis from a G3RS Tamagaki potential (local potential) and
a nonlocal potential which has a simple Gaussian nonlocality
at a short range but produces the same scattering phase shifts 
as the local potential does.
The nonlocal Gaussian soft core is typical of quark cluster model (QCM)
which gives a good description of the nucleon-nucleon scattering.

The G-matrix elements thus derived from the nonlocal potential show
an effective repulsion in comparison with those derived from 
the local potential.  
The difference appears in the SE and TE channels mostly.
The effects are, however, generally very small for all nuclei.   
It is not, therefore, likely that quark-originated nonlocal effects 
are detected in conventional nuclear phenomena.

\pagebreak

\begin{figure}
\epsfxsize 450 pt
\centerline{\epsfbox{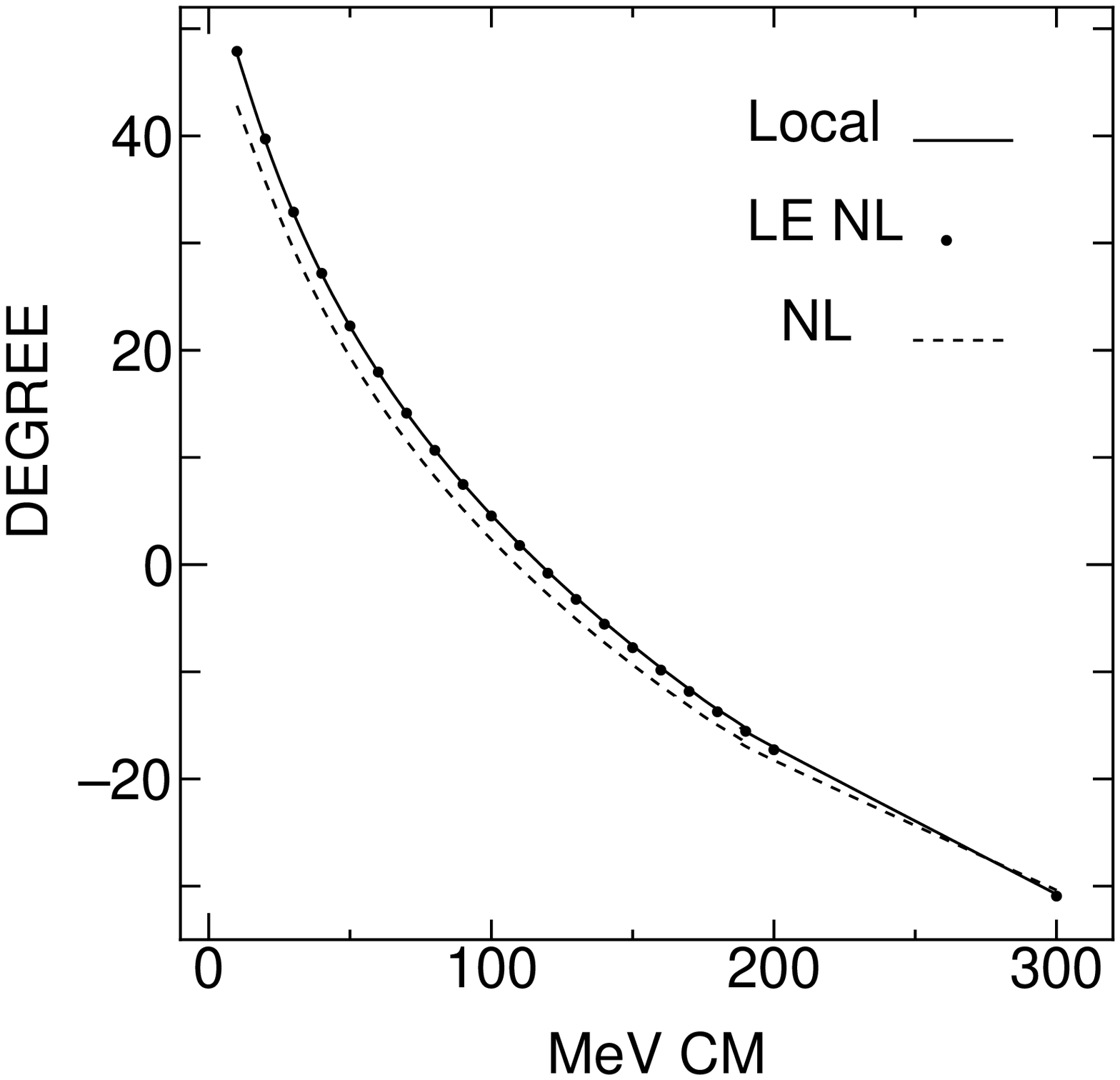}}
\caption{$NN$ scattering phase shifts in the $^1S_0$ channel 
for the local, LENL and NL potentials. }
\end{figure}

\begin{figure}
\epsfxsize 450 pt
\centerline{\epsfbox{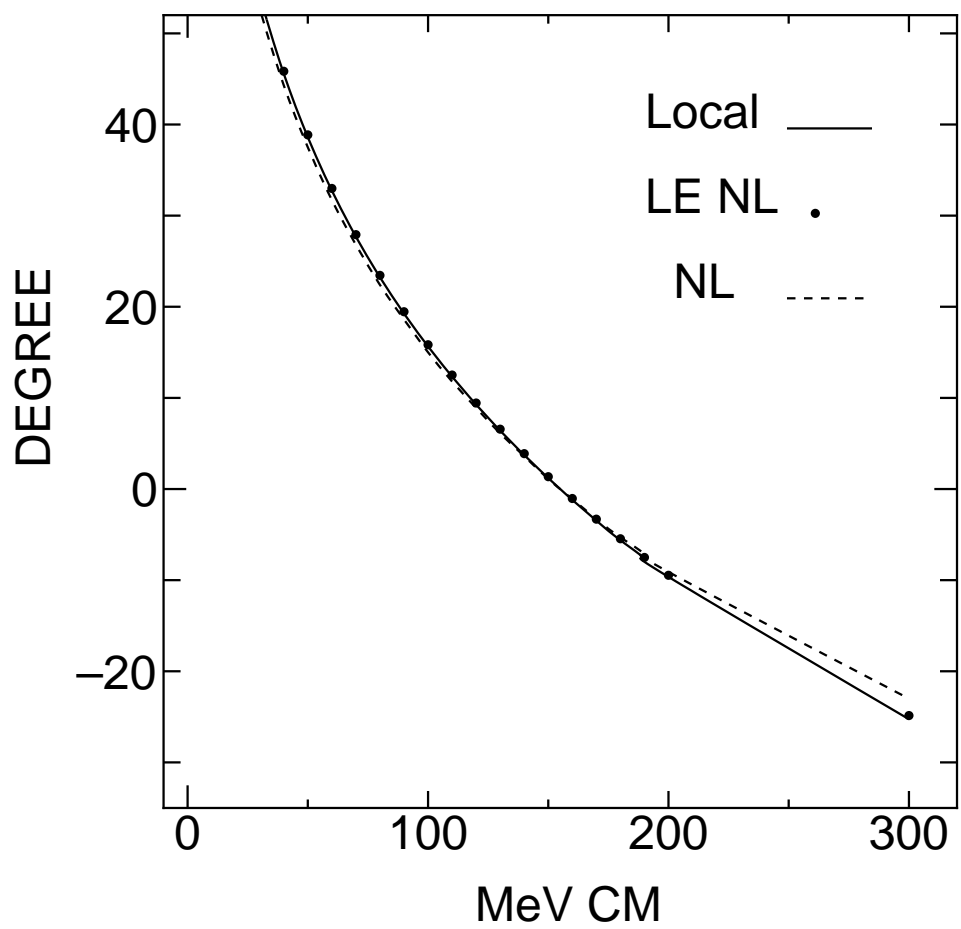}}
\caption{$NN$ scattering phase shifts for $^3S_1$. }
\end{figure}

\end{document}